\documentclass[12pt,a4paper]{article}
\usepackage{graphicx}
\usepackage{lineno}
\usepackage{amsmath}
\usepackage{bm}
\usepackage{natbib}
\usepackage{todonotes}
\usepackage{geometry}
\usepackage{graphicx}
\usepackage{subfigure}

\newcommand{\beq}{\begin{equation}}
\newcommand{\eeq}{\end{equation}}

\newcommand{\bz}{\mathbf{z}}
\newcommand{\ba}{\mathbf{a}}

\newcommand{\bx}{\mathbf{x}}
\newcommand{\br}{\mathbf{r}}
\newcommand{\bg}{\mathbf{g}}
\newcommand{\bs}{\mathbf{s}}
\newcommand{\bX}{\mathbf{X}}

\newcommand{\by}{\mathbf{y}}
\newcommand{\bY}{\mathbf{Y}}

\newcommand{\bG}{\mathbf{G}}

\newcommand{\bE}{\mathbf{E}}

\newcommand{\bA}{\mathbf{A}}
\newcommand{\bJ}{\mathbf{J}}

\newcommand{\bS}{\mathbf{S}}

\newcommand{\bZ}{\mathbf{Z}}
\newcommand{\bD}{\mathbf{D}}
\newcommand{\bI}{\mathbf{I}}

\begin{document}

\title{\textsc{Heterofusion: fusing genomics data of different measurement scales}}

\author{A.K. Smilde*$^{1}$, Y. Song$^1$, J.A. Westerhuis$^1$, H.A.L. Kiers$^2$, \\N. Aben$^3$, L.F.A. Wessels$^3$}

\maketitle

\begin{tabular}{l}
$^1$Biosystems Data Analysis, Swammerdam Institute for Life Sciences, \\
University of Amsterdam, Amsterdam, The Netherlands.\\
$^2$Heymans Institute, University of Groningen, Groningen, The Netherlands.\\
$^3$Oncode Institute, Netherlands Cancer Institute, Amsterdam, The Netherlands.\\

$^*$Corresponding author (a.k.smilde@uva.nl)\\
\end{tabular}

\section{Abstract}

In systems biology, it is becoming increasingly common to measure biochemical entities at different levels of the same biological system. Hence, data fusion problems are abundant in the life sciences. With the availability of a multitude of measuring techniques, one of the central problems is the heterogeneity of the data. In this paper, we discuss a specific form of heterogeneity, namely that of measurements obtained at different measurement scales, such as binary, ordinal, interval and ratio-scaled variables. Three generic fusion approaches are presented of which two are new to the systems biology community. The methods are presented, put in context and illustrated with a real-life genomics example.\\

\noindent Keywords: data fusion, data integration, low-level fusion, measurement scales, heterogeneous data

\section{Introduction}

\subsection{General}

With the availability of comprehensive measurements collected in multiple related data sets in the life sciences, the need for a simultaneous analysis of such data to arrive at a global view on the system under study is of increasing importance. There are many ways to perform such a simultaneous analysis and these go also under very different names in different areas of data analysis: data fusion, data integration, global analysis, multi-set or multi-block analysis to name a few. We will use the term \emph{data fusion} in this paper.\\

\noindent Data fusion plays an especially important role in the life sciences, e.g., in genomics it is not uncommon to measure gene-expression (array data or RNA-sequencing (RNAseq) data), methylation of DNA and copy number variation. Sometimes, also proteomics and metabolomics measurements are available. All these examples serve to show that having methods in place to integrate these data is not a luxury anymore.

\subsection{Types of data fusion}

Without trying to build a rigorous taxonomy of data fusion it is worthwhile to distinguish several distinctions in data fusion. The first distinction is between model-based and exploratory data fusion. The former uses background knowledge in the form of models to fuse the data; one example being genome-scale models in biotechnology \citep{Zimmermann2017}. The latter does not rely on models, since these are not available or poorly known, and thus uses empirical modeling to explore the data. In this paper, we will focus on exploratory data fusion.\\

\noindent The next distinction is between low-, medium-, and high-level fusion \cite{Steinmetz1999}. In low-level fusion, the data sets are combined at the lowest level, that is, at the level of the (preprocessed) measurements. In medium-level fusion, each separate data set is first summarized, e.g., by using a dimension reduction method or through variable selection. The reduced data sets are subsequently subjected to the fusion. In high-level fusion, each data set is used for prediction or classification of an outcome and the prediction or classification results are then combined, e.g, by using majority voting \citep{Doeswijk2011}. In machine learning this is known as early, intermediate and late integration. All these types of fusions have advantage and disadvantages which are beyond the scope of this paper. In this paper, we will focus on low- and medium-level fusion.\\

\noindent The final characteristic of data fusion relevant for this paper is heterogeneity of the data sets to be fused. Different types of heterogeneity can be distinguished. The first one is the type of \emph{data}, such as metabolomics, proteomics and RNAseq data in genomics. Clearly, these data relate to different parts of the biological system. The second one is the type of \emph{measurement-scale} in which the data are measured that are hoing to be fused. In genomics, an example is a data set where gene-expressions are available and mutation data in the processed form of Single Nucleotide Variants (SNVs). The latter are binary data and gene-expression is quantitative data. They are clearly measured at a different scale. Ideally, data fusion methods should consider the scale of such measurements and this will be the topic of this paper.

\subsection{Types of measurement scales}

The history of measurement scales goes back a long time. A seminal paper drawing attention to this issue appeared in the 40-ties \citep{Stevens1946}. Since then numerous papers, reports and books have appeared \citep{Suppes1962,Krantz1971,Narens1981,Narens1986,Luce1987,Hand1996}. The measuring process assigns numbers to aspects of objects (an \textit{empirical system}), e.g, weights of persons. Hence, measurements can be regarded as a mapping from the empirical system to numbers, and scales are properties of these mappings.\\

\noindent In measurement theory, there are two fundamental theorems \citep{Krantz1971}: the representation theorem and the uniqueness theorem. The \textit{representation theorem} asserts the axioms to be imposed on an empirical system to allow for a homomorphism of that system to a set of numerical values. Such a homomorphism into the set of real numbers is called a scale and thus represents the empirical system. A scale possesses \textit{uniqueness} properties: we can measure the weight of persons in kilograms or in grams, but if one person weighs twice as much as another person, this ratio holds true regardless the measurement unit. Hence, weight is a so-called ratio-scaled variable and this ratio is unique. The transformation of measuring in grams instead of kilograms is called a \textit{permissible} transformation since it does not change the ratio of two weights. For a ratio-scaled variable, only similarity transformations are permissible; i.e. $\widetilde{x}=\alpha x; \alpha>0$ where $x$ is the variable on the original scale and $\widetilde{x}$ is the variable on the transformed scale. This is because
\begin{equation}\label{eRatio}
 \frac{\widetilde{x_i}}{\widetilde{x_j}}=\frac{\alpha x_i}{\alpha x_j}=\frac{x_i}{x_j}.
\end{equation}
Note that this coincides with the intuition that the unit of measurement is immaterial.\\

\noindent The next level of scale is the interval-scaled measurement. The typical example of such a scale is degrees Celsius and the permissible transformation is affine; i.e. $\widetilde{x}=\alpha x +\beta; \alpha>0$. In that case, the ratio of two intervals is unique because
\begin{equation}\label{eInterval}
 \frac{\widetilde{x_i}-\widetilde{x_j}}{\widetilde{x_k}-\widetilde{x_l}}=\frac{(\alpha x_i + \beta)-(\alpha x_j + \beta)}{(\alpha x_k + \beta)-(\alpha x_l + \beta)}=\frac{\alpha (x_i-x_j)}{\alpha (x_k-x_l)}=\frac{x_i-x_j}{x_k-x_l}.
\end{equation}
Stated differently, the zero point and the unit are arbitrary on this scale.\\

\noindent Ordinal-scaled variables represent the next level of measurements. Typical examples are scales of agreement in surveys: strongly disagree, disagree, neutral, agree and strongly agree. There is a rank-order in these answers, but no relationship in terms of ratios or intervals. The permissible transformation of an ordinal-scale is a monotonic increasing transformation since such transformations keep the order of the original scale intact.\\

\noindent Nominal-scaled variables are next on the list. These variables are used to encode categories and are sometimes also called categorical. Typical example are gender, race, brands of cars and the like. The only permissible transformation for a nominal-scaled variable is the one-to-one mapping. A special case of a nominal-scaled variable is the binary (0/1) scale. Binary data can have different meanings; they can be used as categories (e.g. gender) and are then nominal-scale variables. They can also be two points on a higher-level scale, such as absence and presence (e.g. for methylation data).\\

\noindent The above four scales are the most used ones but others exists \citep{Suppes1962,Krantz1971}. Counts, e.g., have a fixed unit and are therefore sometimes called absolute-scaled variables \citep{Narens1986}. Another scale is the one for which the power transformation is permissible; i.e. $\widetilde{x}=\alpha x^\beta; \alpha, \beta>0$ which is called a log-interval scale because a logarithmic transformation of such a scale results in an interval-scale. An example is density \citep{Krantz1971}. Sometimes the scales are lumped in quantitative (i.e. ratio and interval) and qualitative (ordinal and nominal) data.\\

\noindent An interesting aspect of measurement scales is to what extent meaningful statistics can be derived from such scales (see Table 1 in \citep{Stevens1946}). A prototypical example is using a mean of a sample of nominal-scaled variables which is generally regarded as being meaningless. This has also provoked a lot of discussion \citep{Adams1965,Hand1996} and there are nice counter-examples of apparently meaningless statistics that still convey information about the empirical system \citep{Michell1986}. As always, the world is not black or white.

\subsection{Motivating example}

Examples of fusing data of different measurement scales are abundant in modern life science research. We will first give a short description of modern measurements in genomics that will illustrate this. In a sample extracted from biological systems (e.g. cells) it is possible to measure the mRNA molecules. This is done nowadays with RNAseq techniques and in essence the mRNA are counts per volume, hence, a concentration. Epigenetics concerns, amongst other, the methylation of some of the sites of a DNA molecule and is in essence a binary variable (yes/no methylated at a given location of the DNA). Another feature in genetics is whether a location of the DNA is mutated, a phenomenon called SNVs (single nucleotide variants), which is also binary. Lastly, there are Copy Number Variations (CNVs) of genes on the genome which is in essence a (limited) number of counts and sometimes expressed as Copy Number Abberations (CNA) with a binary coding (no: normal number of copies, yes: aberrant number of copies). If we move to the field of metabolomics and proteomics, then most of the measurements are relative intensities and in some cases - when calibration lines have been made - concentrations which are ratio-scaled.\\

\noindent The above exposition clearly shows that if we want to fuse different types of genomics data, or fuse genomics data with metabolomics and/or proteomics then there is a problem of different measurement scales. This problem is aggravated by the fact that some of this data is very high-dimensional. SNP and methylation data can contain 100.000 features or variables, RNAseq data has usually around 20.000 genes. Shotgun proteomics data (based on LC-MS or LC-MS/MS) can also easily contain 100.000 features. Hence, in many cases dimension reduction has to take place, asking for methods to deal properly with the corresponding measurement scale. For some of the methods to be discussed in this paper there are already examples in the literature. There are examples of the use of the parametric approach using latent variables \citep{Shen2009,Mo2013} and also of the optimal scaling approach \citep{Wietmarschen2011,Wietmarschen2012}. For the third approach to be discussed, we have not found examples yet in the life sciences. We will come back to these examples in Section \ref{Discussion}.

\subsection{Goal of the paper}

In this paper, we describe low- and mid-level fusion ideas of data of different measurement scales. We will restrict ourselves to data sharing the object mode. Mid-level fusion first selects variables and then is subjected to the methods described below. These methods can be applied in different fields of science, but we will illustrate them by using a genomics example.\\

\noindent We think this paper is needed since the different methods originate from different fields of data analysis, psychometrics and bioinformatics with limited cross-talk between those fields; we will try to fill this gap. Moreover, there are relationships between the methods and this might help in selecting the proper method for a particular application. Hence, we will also discuss the properties of the different methods.\\

\noindent We will select and discuss methods that provide coordinates of the objects that can be used for plotting and visualizing the relationships between the objects. Moreover, we think it is also worthwhile to consider methods that generate importance values for the variables in the different blocks since this will facilitate interpretation of the results in substantive terms.

\section{Theory}

\subsection{Three basic ideas}

We will describe three basic ideas that can be used for fusion of data of different measurement scales on a conceptual level. A more detailed explanation is given in following Sections. One of these methods is parametric and thus depends on distributions \citep{Mo2013}. The other two methods are non-parametric and based on concepts of representation matrices \citep{Zegers1986c,Kiers1989} and optimal scaling \citep{Gifi1990}.\\

\begin{figure}[h!]
 \centerline{\includegraphics*[width=10cm]{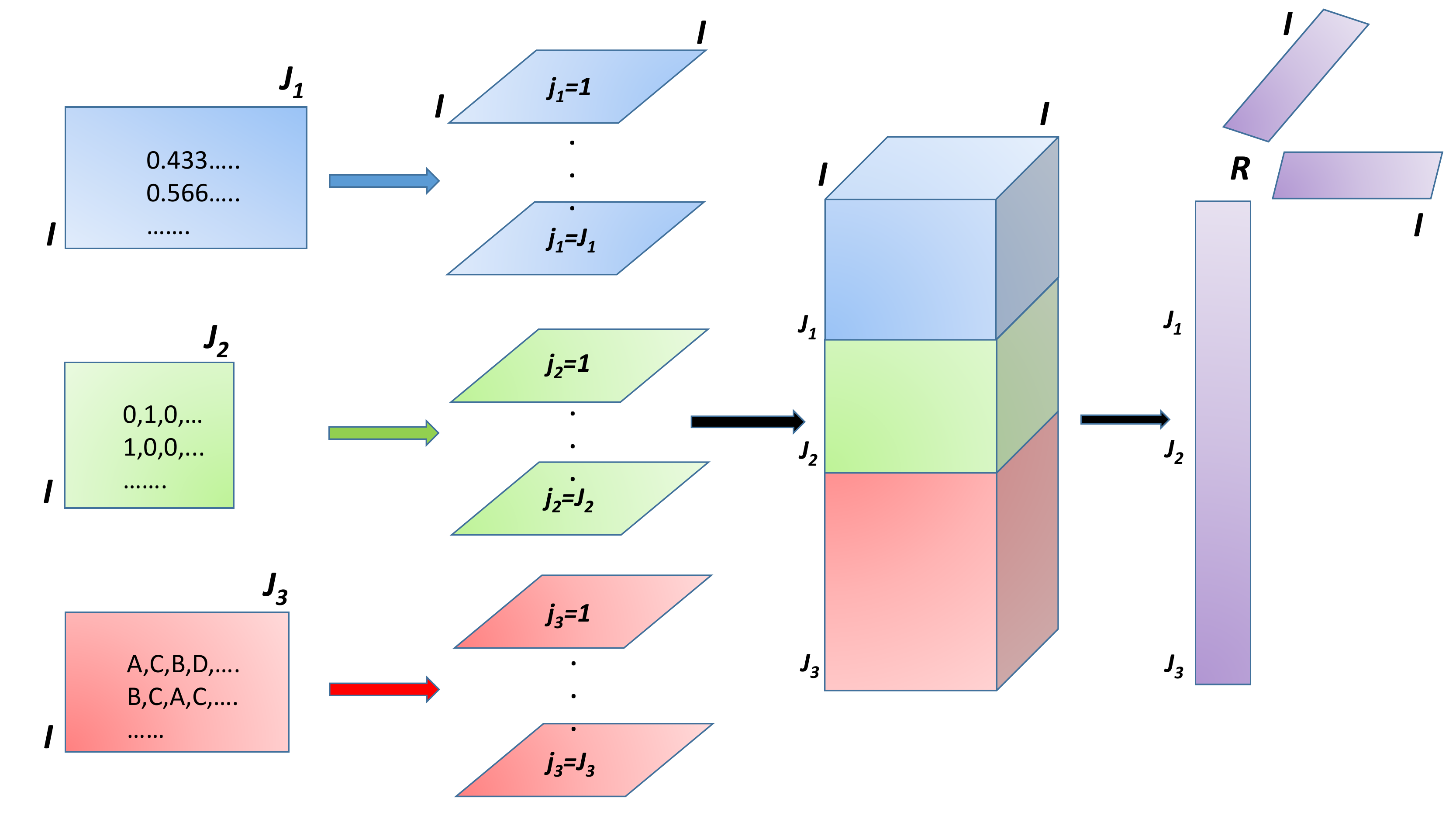}}
 \caption{\footnotesize Heterofusion using representation matrices (see text).}
 \label{FigureRM}
\end{figure}

\noindent The first idea is illustrated in Figure \ref{FigureRM} \citep{Kiers1989}. Suppose we have three blocks of data, the first block ($\bX_1$) contains ratio-scaled data, the second block ($\bX_2$) binary data and the third block ($\bX_3$) categorical data with each of the $J_3$ variables having four categories (labeled A, B, C and D). Each variable in each block is represented by an $I \times I$ representation matrix (to be explained later). Then these representation matrices can be stacked and the resulting three-way array can be analyzed by a suitable three-way method using $R$ components giving coordinates for the objects and weights for the variables.\\

\begin{figure}[h!]
 \centerline{\includegraphics*[width=10cm]{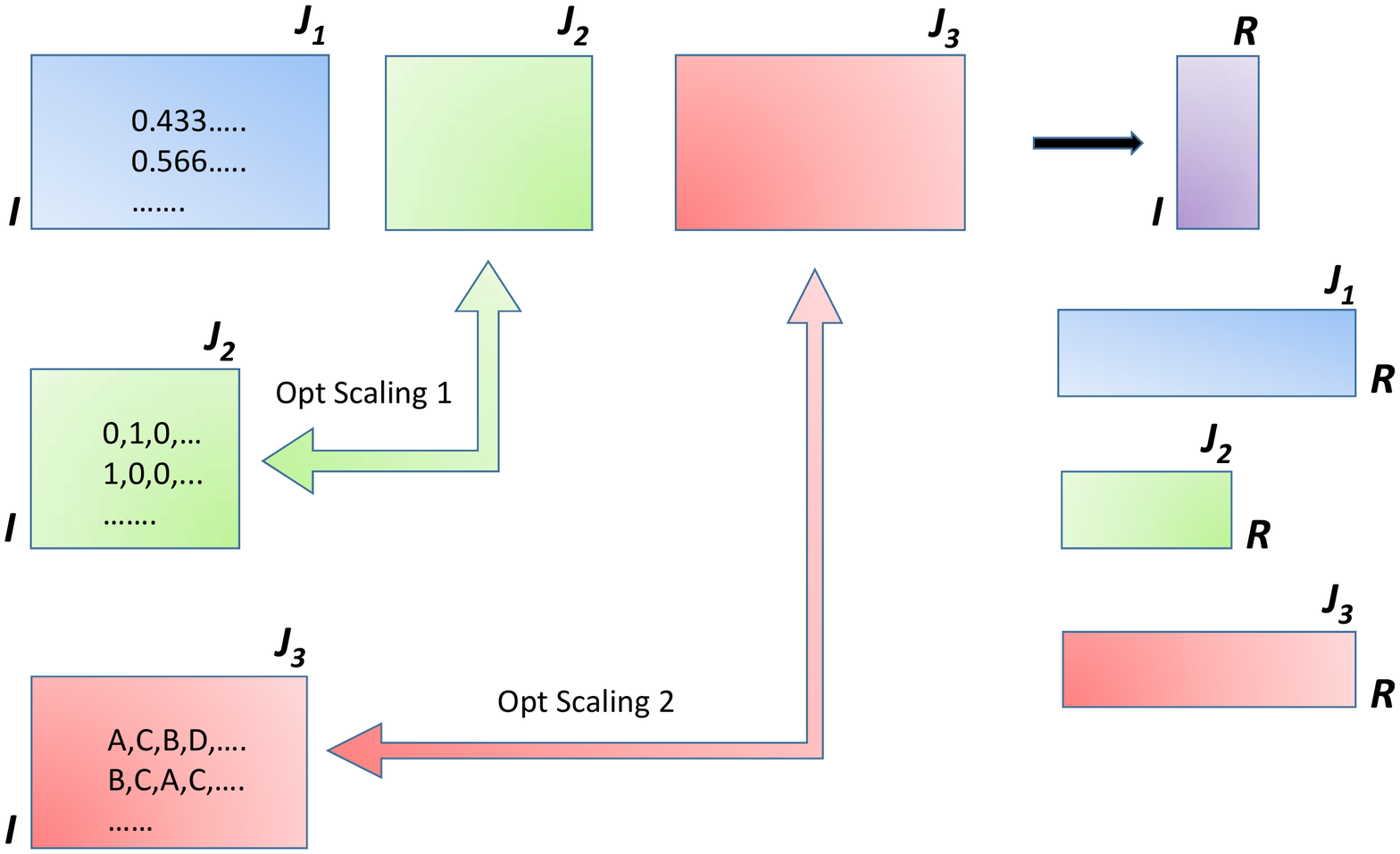}}
 \caption{\footnotesize Heterofusion using optimal scaling (see text).}
 \label{FigureOS}
\end{figure}

\noindent The second idea is illustrated in Figure \ref{FigureOS} \citep{Gifi1990,Michailidis1998}. The original matrices are subjected to optimal scaling and the fusion problem is solved as one global optimization problem (to be explained later). The idea of optimal scaling goes back already to R. Fisher and nice introductions are available \citep{Young1981}. For the first block, the variables remain the same but for the second and third block these variables are (optimally) transformed. Using optimal scaling, the three blocks are made comparable and are analyzed simultaneously by a multiblock method (e.g. Simultaneous Component Analysis or Consensus PCA) giving $R$ coordinates for the objects (the $I \times R$ matrix) and loadings (the ($J_1 \times R$), ($J_2 \times R$) and ($J_3 \times R$) matrices) for the transformed variables.\\

\begin{figure}[h!]
 \centerline{\includegraphics*[width=10cm]{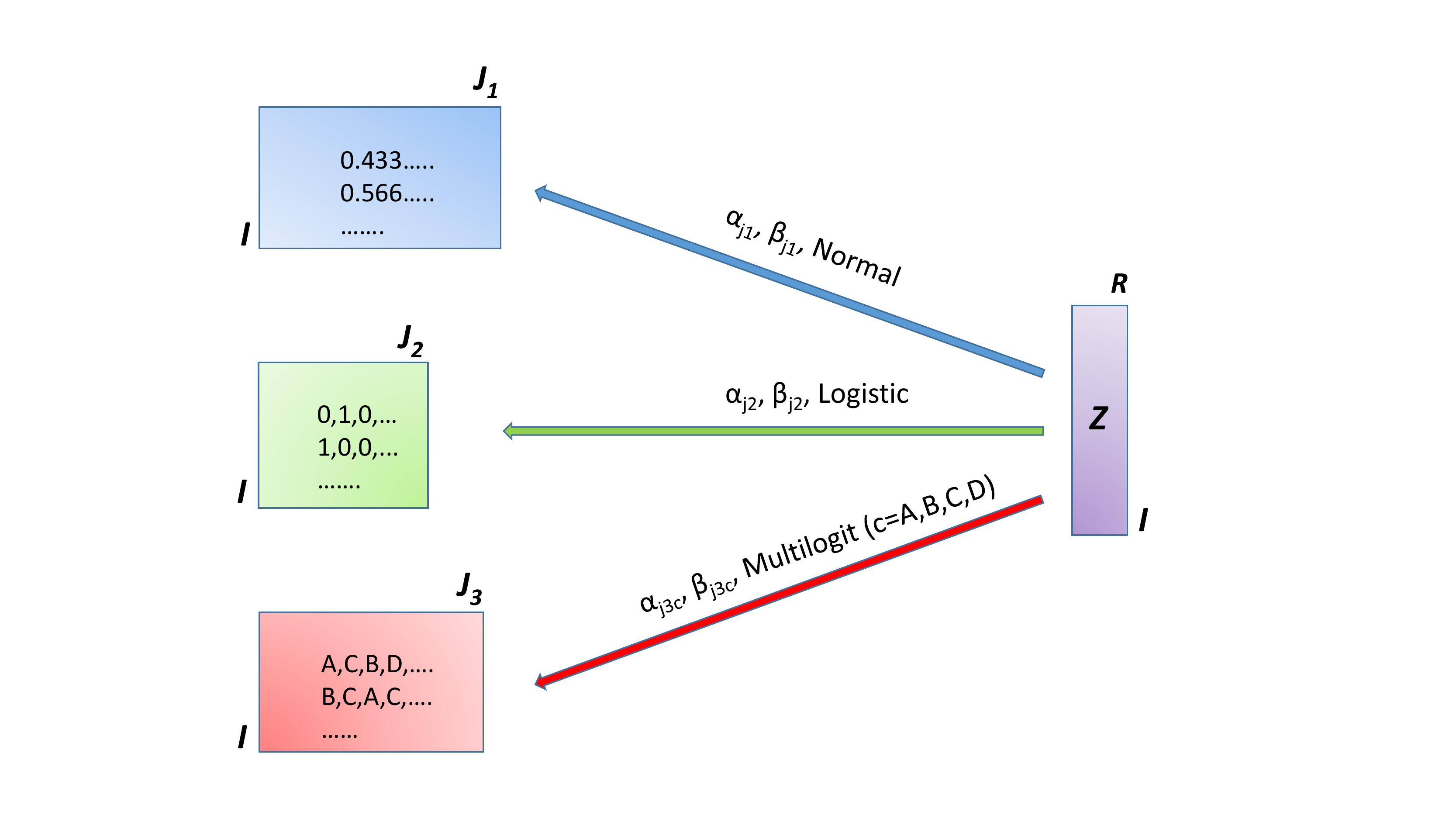}}
 \caption{\footnotesize Heterofusion using parametric models (see text).}
 \label{FigurePM}
\end{figure}

\noindent The third idea relies on the explicit use of the $R$ latent variables collected in $\bZ$ (see Figure \ref{FigurePM}) \citep{Mo2013}. These latent variables are then thought to generate the manifest variables in the different blocks using different distributions. For the ratio-scaled block, a regression model is assumed based on the normal distribution and with parameters $\boldsymbol{\alpha_{j1}}$ and $\boldsymbol{\beta_{j1}}$. For the binary block, a logit or probit model is assumed with parameters $\boldsymbol{\alpha_{j2}}$ and $\boldsymbol{\beta_{j2}}$. The final - categorical - block is modeled by a multilogit model with parameters $\boldsymbol{\alpha_{j3c}}$ and $\boldsymbol{\beta_{j3c}}$ where $c=A,B,C,D$.\\

\noindent We will use the following conventions for notations. A vector $\bx$ is a bold lowercase and a matrix ($\bX$) a bold uppercase. Running indices will be used for samples ($i$=1,\ldots ,$I$) with $I$ is the number of samples; we will use likewise the indices $k$=1,\ldots ,$K$ for the data blocks; variables within a data block are indexed by $j_k$=1,\ldots ,$J_k$ and we will use $r$=1,\ldots ,$R$ as an index for latent variables or components.\\

\subsection{Representation matrices approaches}

\subsubsection{Representation matrices}

\noindent \textsc{Idea of representation matrices.}\\

\noindent Suppose we have a data matrix $\bX (I \times J)$ with columns $\bx_j$ containing the scores of the objects on variable $j$. Such a score can be a ratio-scaled value, but can also be a binary value, a categorical value or an ordinal-scaled value. A representation operator works on this vector and produces a representation matrix which serves as a building block to calculate associations between variables and to analyze several variables simultaneously \citep{Zegers1986c, Kiers1989}. Such a representation matrix can be a vector ($I \times 1$), a rectangular matrix ($I \times R; R<I$) or a square matrix ($I \times I$). Let $\bS_j$ and $\bS_k$  be the representation matrices for variables $j$ and $k$, respectively, then a general equation of the association between variables $j$ and $k$ is
\begin{equation}\label{eAss1}
    q_{jk}=\frac{2tr(\bS_j^T\bS_k)}{tr(\bS_j^T\bS_j)+tr(\bS_k^T\bS_k)}
\end{equation}
where the symbol 'tr' is used to indicate the trace of a matrix. In most cases that follow below the representation matrices are standardized (centered and scaled to length one\footnote{An alternative is scaling to variance one, but this only differs with the same constant for each variable.}) and in these cases Eqn. \ref{eAss1} simplifies to
\begin{equation}\label{eAss2}
    \tilde{q}_{jk}=tr(\bS_j^T\bS_k)
\end{equation}
since both $tr(\bS_j^T\bS_j)$ and $tr(\bS_k^T\bS_k)$ are one. As will be shown in the following, Eqn. \ref{eAss2} can generate the familiar associations such as the Pearson correlation or the Spearman correlation. An extensive description of all kinds of representation matrices is beyond the scope of this paper; we will discuss the most relevant ones for the problem of heterofusion. The idea of representation matrices\footnote{Their original name was quantification matrices but that name has also been used differently. Hence our choice to rename such matrices.} goes back to the work of \citet{Janson1982} and \citet{Zegers1986c}. Examples of different representation matrices are given in Section \ref{Appendix}.\\

\noindent \textsc{Representation matrices for ratio- and interval-scaled values.}\\

\noindent For ratio- and interval-scaled values, two types of representation matrices can be defined: vectors and square matrices. If $\bx_j$ represents the raw scores of the objects on variable $j$ then the vector quantification can be this vector itself (i.e. $\bs_j=\bx_j$) or a standardized version of it. When the latter is used in Eqn. \ref{eAss2}, Pearson's $R$-value is obtained. In standard multivariate analysis this is by far the most used representation matrix.\\

\noindent There is also another possibility for ratio- and interval-scaled values, namely square representation matrices. Two examples are the following. Define
\begin{equation}\label{eRM1}
    \widetilde{\bS}_j=(\bx_j\textbf{1}^T-\textbf{1}\bx_j^T)
\end{equation}
where $\textbf{1}$ is an $I \times 1$ column of ones. This $\widetilde{\bS}_j$ generates a skew-symmetric matrix enumerating all differences between the object-scores of variable $j$ (for an example, see \ref{Examples ratio scaled}). Hence, distances between objects are obtained per variable and these distance matrices can be subjected to an INDSCAL model \citep{Kiers1989}. Upon standardizing $\widetilde{\bS}_j$ by $\bS_j=(tr\widetilde{\bS}_j^T\widetilde{\bS}_j)^{-1/2}\widetilde{\bS}_j$ and using this $\bS_j$ (and a similarly defined $\bS_k$) in Eqn. \ref{eAss2} gives again Pearson's $R$-value. Another example is using $\bS_j=\bs_j\bs_j^T$ where $\bs_j$ is the standardized version of $\bx_j$. Using this $\bS_j$ (and a similarly defined $\bS_k$) in Eqn. \ref{eAss2} gives Pearson's $R^2$ value. Such representation matrices correspond to the blue-squared matrices in Figure \ref{FigureRM} and are the basis of Kernel and Multidimensional Scaling methods (\textbf{check!}).\\

\noindent \textsc{Representation matrices for ordinal-scaled values.}\\

\noindent When the data are ordinal-scaled, then the vector of readings can be encoded in terms of rank-orders $\br_j (I \times 1)$. For the earlier example of strongly disagree, disagree, neutral, agree, strongly agree such a ranking may be encoded as 1 (strongly disagree) to 5 (strongly agree). Then again - as in the ratio-scaled variables - representation can be done using the vectors $\br_j$ or their standardized version. In the latter case, applying Eqn. \ref{eAss2} to this version gives the Spearman's rank-order correlation coefficient. Another representation is by using (the raw-)$\br_j$ in Eqn. \ref{eRM1} instead of $\bx_j$ and this generates Spearman's rank-order correlation coefficient after using Eqn. \ref{eAss1}.\\

\noindent \textsc{Representation matrices for nominal-scaled values.}\\

\noindent We will discuss the representation matrices for nominal-scaled variables separately for binary data and categorical data. We first discuss representation matrices for categorical data. We have to distinguish two situations: one in which all categorical variables have the same number of categories and the situation that this is not the case. Since the latter is more general and encountered more often, we will restrict ourselves to this case. Then only square representation matrices are available. These are based on indicator matrices \citep{Zegers1986c,Kiers1989,Gifi1990}. If variable $\bx_j$ has four categories (A,B,C,D), then this can be encoded in the rectangular matrix $\bG_j (I \times 4)$ where each column $\bg_{jk}$ in $\bG_j$ represents a category and each row an object. This matrix has only zeros or ones; where $g_{ijk}$ is one, if and only if object $i$ belongs to the category represented by $k$. The representation matrix can now be built using the products $\bG_j \bG_j^T (I \times I)$. There are very many versions of such square representation matrices based on indicator matrices and some of them give rise to a known correlation, e.g.,
\begin{equation}\label{eRMNom1}
    \bJ \bG_j \bD_j^{-1} \bG_j^T \bJ
\end{equation}
where $\bJ (I \times I)$ is the centering operator and $\bD_j (C_j \times C_j)$ is a diagonal matrix containing the marginal frequencies of categories $1,..,C_j$ for variable $j$. The corresponding correlation coefficient is the so-called $T^2$ coefficient \citep{Tschuprow1939}. These representation matrices correspond to the red-square matrices in Figure \ref{FigureRM}. Examples are given in Section \ref{Examples nominal data}.\\

\noindent For binary data (if all variables are binary) then rectangular representation matrices are possible. This comes down to the same idea as above, namely, to consider the binary variables as representing two categories. This results then in representation matrices $\bG_j$ of sizes $(I \times 2)$ . When fusing with other types of variables is the goal, then a squared type of representation is needed such as in Eqn. \ref{eRMNom1} and visualized in Figure \ref{FigureRM} (green matrices). Examples are given in Section \ref{Examples binary data}.\\

\subsubsection{Data fusion using representation matrices}
\label{Data fusion using representation matrices}

To illustrate how to use representation matrices we will work with four data matrices, each on a different measurement scale and sharing the same set of $I$ samples. The first matrix $\bX_1 (I \times J_1)$ contains ratio- or interval-scaled data; the second matrix $\bX_2 (I \times J_2)$ contains ordinal-scaled data; the third $\bX_3 (I \times J_3)$ contains nominal data and the last matrix $\bX_4 (I \times J_4)$ contains binary data.\\

\noindent The representation matrices $\bS_j$ can now be used in a three-way model for symmetric data. The basic model for a single data block is the INDSCAL (INdividual Differences SCALing) model:
\begin{equation}\label{eINDSCAL}
    \min_{\bZ,\bA_j} \sum_{j=1}^J||\bS_j-\bZ \bA_j \bZ^T||^2
\end{equation}
where $\bA_j$ is the diagonal matrix with the $j^{th}$ row of the loadings $\bA (J \times R)$ on its diagonal and the matrix $\bZ(I \times R)$ contains the object scores. The loadings $\bA (J \times R)$ are nonnegative to ensure the fitted part of the model ($\bZ \bA_j \bZ^T$) to be positive (semi-) definite.  If the additional constraint that $\bZ^T\bZ=\bI$ is used, then the model is called INDORT (INDscal with ORThogonal constraints) \citep{Kiers1989}.\\

\noindent The INDORT method can now be generalized to analyze simultaneously all blocks by simply stacking all similarity matrices on top of each other (see Figure \ref{FigureRM}):
\begin{equation}\label{eINDOMIX}
    \min_{\bZ,\bA_{j_k}} \sum_{k=1}^4 \sum_{j_{k}=1}^{J_{k}}||\bS_{j_k}-\bZ \bA_{j_k} \bZ^T||^2
\end{equation}
where $\bA_{j_k}$ is the diagonal matrix with the $j_k^{th}$ row of the loadings $\bA_k (J \times R)$ on its diagonal and the matrix $\bZ(I \times R)$ contains the object scores. This model is called IDIOMIX for obvious reasons \citep{Kiers1989}.

\subsection{Optimal scaling approaches}

There are many ways to explain optimal scaling; we will follow the exposition given by \cite{Michailidis1998}. Suppose that the matrix $\bX (I \times J)$ contains $J$ categorical variables not necessarily with the same number of categories. Each variable $\bx_j$ can now be encoded with indicator matrix $\bG_j (I \times L_j)$ where $L_j$ is the number of categories for variable $j$ as discussed before. The idea of optimal scaling is to find objects scores $\bZ (I \times R)$ and category quantification matrices $\bY_j (L_j \times R; j=1,...,J)$ such that the following problem is solved \citep{Michailidis1998}:
\begin{equation}\label{eOS1}
    \min_{\bZ,\bY_j} \sum_{j=1}^J||\bZ-\bG_j \bY_j||^2
\end{equation}
under the constraints that $(1/I) \bZ^T \bZ = \bI$ and these scores are centered around zero (to avoid trivial solutions of Eqn. \ref{eOS1}). This method - including the alternating optimization method to solve Eqn. \ref{eOS1} - is called homogeneity analysis or HOMALS for short \citep{Gifi1990}. The rows of $\bZ$ give a low dimensional representation of the objects and the matrices $\bY_j (j=1,...,J)$ give the optimal quantifications of the categorical variables. Note that these matrices $\bY_j (j=1,...,J)$ are not loadings; they give quantifications for the categorical variables which are different for the $R$ components, namely $\by_{jr} (L_j \times 1; r=1,...,R)$ where $\by_{jr}$ is the $r-th$ column of $\bY_j$.\\

\noindent Upon restricting the rank of $\bY_j (j=1,...,J)$ to be one, we arrive at non-linear PCA (PRINCALS) \citep{Gifi1990,Michailidis1998}. Then Eqn. \ref{eOS1} can be rewritten as
\begin{equation}\label{eOS2}
    \min_{\bZ,\by_j,\ba_j} \sum_{j=1}^J||\bZ-\bG_j \by_j {\ba^T}\!_j||^2
\end{equation}
with the same constraints on $\bZ$ as before (i.e. $(1/I)\bZ^T\bZ=\bI$). As an identification constraint for $\by_j$ and $\ba_j$ we impose $\by_j^T\bG_j^T\bG_j\by_j=I$. Now, the vectors $\ba_j (R \times 1)$ are the loadings and $\by_j (L_j \times 1)$ contain the quantifications which are the same for all $R$ dimensions of the solution. The relationship between (linear) PCA and non-linear PCA becomes clear when rewriting Eqn. \ref{eOS2} (following \citep{Gifi1990}, p.167-168) as
\begin{gather}\label{eOS3}
    \min_{\bZ,\by_j,\ba_j} \sum_{j=1}^J||\bZ-\bG_j \by_j \ba_j^T||^2= \\ \nonumber
    \min_{\bZ,\by_j,\ba_j} \sum_j tr(\bZ^T\bZ)-2 \sum_j tr(\bZ^T\bG_j\by_j\ba_j^T)+\sum_jtr(\ba_j\by_j^T\bG_j^T\bG_j\by_j\ba_j^T)= \\ \nonumber
    \min_{\bZ,\by_j,\ba_j} IJtr\bI-2\sum_j tr(\bZ^T\bG_j\by_j\ba_j^T)+I \sum_j tr(\ba_j \ba_j^T) \nonumber
\end{gather}
using the constraints on $\bZ$ and $\by_j$. The function in Eqn. \ref{eOS3} differs only a constant from the function
\begin{equation}\label{eOS4}
    g(\bZ,\by_j,\ba_j)=\sum_j\|\bG_j\by_j-\bZ\ba_j\|^2,
\end{equation}
as follows from rewriting $g(\bZ,\by_j,\ba_j)$ using the constraints on $\bZ$ and $\by_j$:
\begin{gather}\label{eOS5}
    g(\bZ,\by_j,\ba_j)= \\ \nonumber
    \sum_j \by_j^T \bG_j^T \bG_j \by_j-2 \sum_j tr(\ba_j^T \bZ^T \bG_j \by_j)+\sum_j tr(\ba_j^T \bZ^T \bZ \ba_j)= \\ \nonumber
    IJ-2\sum_j tr(\bZ^T \bG_j \by_j \ba_j^T)+I\sum_j tr(\ba_j^T \ba_j). \nonumber
\end{gather}
Thus, it has been shown that problem Eqn. \ref{eOS2} subject to the constraints $(1/I)\bZ^T\bZ=\bI$ and $\by_j^T\bG_j^T\bG_j\by_j=I$ is equivalent to the problem
\begin{gather}\label{eOS6}
    \min_{\bZ,\by_j,\ba_j} \sum_{j=1}^J\|\bG_j \by_j - \bZ_j \ba_j\|^2= \\ \nonumber
    \min_{\bZ,\by_j,\ba_j} \|[\bG_1 \by_1|...|\bG_J \by_J]-\bZ \bA^T\|^2= \\ \nonumber
    \min_{\bZ,\by_j,\ba_j} \|\bX^*-\bZ \bA^T\|^2 \nonumber
\end{gather}
where $\bA$ has rows $\ba_j^T$ and $[\bG_1 \by_1|...|\bG_J \by_J]$ is written as $\bX^*$ where the superscript '*' represents the optimal scaled data, and this is seen to be the (non-linear) analog of ordinary PCA \citep{Gifi1990}.\\

\noindent The nature of the measurement scale can now be incorporated by allowing the quantifications to be free for nominal-scale data and monotonic for ordinal-scaled data. The latter quantification ensures the order in the ordinal-scaled data. Framed in terms of Eqn. \ref{eOS6} this becomes:
\begin{equation}\label{eOS7}
    {x^*}\!_{ij}>{x^*}\!_{kj}\;\; if \;\; x_{ij}>x_{kj}
\end{equation}
where ${x^*}\!_{ij}$ and ${x^*}\!_{kj}$ are elements of $\bX^*$; $x_{ij}$ and $x_{kj}$ are elements of $\bX$. Ties in the original data can be treated in different ways depending on whether the underlying measurements can be considered continuous or discrete \citep{DeLeeuw1976,Takane1977,Young1978} but this is beyond the scope of this paper.\\

\noindent There are close similarities between optimal scaling and multiple correspondence analysis \citep{Kiers1989,Michailidis1998}. Binary data represents a special case. When considered as categorical data, non-linear PCA using optimal scaling is the same as performing a (linear) PCA on the standardized binary data, for a proof, see Appendix \citep{DeLeeuw1973,Kiers1989}.

\subsubsection{Data fusion using optimal scaling matrices}

There are different ways to use optimal scaling for fusing data. One method generalizes (generalized) canonical correlation analysis (OVERALS \citep{VanderBurg1988}) and the other method generalizes simultaneous component analysis (SCA) (MORALS \citep{Young1981}). Experiences with generalized canonical correlations show that this method tends to overfit for high-dimensional data. An attempt to overcome this problem is by introducing sparsity constraints \citep{Waaijenborg2008}, but it is not trivial to combine this with optimal scaling. Hence, we chose to use the extension of SCA. Note that SCA was originally developed for analyzing multiple data sets sharing the same set of variables \citep{TenBerge1992}, but it can likewise be formulated for multiple data sets having the sampling mode in common \citep{VanDEun2009}. Using the latter interpretation of SCA leads to the following approach.

We take the same data matrices as in Section \ref{Data fusion using representation matrices} and upon writing $\bX^*=[{\bX^*}\!_1|{\bX^*}\!_2|{\bX^*}\!_3|{\bX^*}\!_4]$ the problem becomes
\begin{equation}\label{eOS8}
    \min_{Par}||\bX^*-\bZ \bA^T||^2=\min_{Par}||[{\bX^*}\!_1|{\bX^*}\!_2|{\bX^*}\!_3|{\bX^*}\!_4]-\bZ [{\bA^T}\!_1|{\bA^T}\!_2|{\bA^T}\!_3|{\bA^T}\!_4]||^2
\end{equation}
with an obvious partition of the loading matrix $\bA$ and where the term 'Par' stands for all parameters. Apart from the scores $\bZ$ and loadings $\bA$ these are the following. For the ratio- interval-scaled block there are no extra parameters since the original scale is used (i.e. ${\bX^*}\!_1=\bX_1$. The second - ordinal-scaled - block puts restrictions on ${\bX^*}$ following the restrictions of Eqn. \ref{eOS7}. The third (nominal-) block has underlying indicator matrices $\bG$ and associated quantifications $\by$ and loadings $\bA_3$ obey the rules of Eqn. \ref{eOS2}. Finally, the binary block ${\bX^*}\!_4$ is simply the standardized version of $\bX_4$ and this ensures an optimal scaling as mentioned above. Note that this way of fusing data assumes an identity link function \citep{VanMechelen2010} and is thus an extension of methods like Consensus PCA and SCA. We will call this method OS-SCA in the sequel. There is no differentiation between common and distinct components \citep{Smilde2017}\\

\subsection{Parametric approaches}

A different class of methods has its roots in factor analysis and can be summarized as follows (see Figure \ref{FigurePM}). The basic idea is that a set of (shared) latent variables is responsible for the variation in all the blocks \citep{Shen2009,Curtis2012,Mo2013} and, subsequently, models are built for the individual blocks based on those shared latent variables.  We will describe the Generalized Simultaneous Component Analysis (GSCA) method \citep{Song2018} in more detail since that is the method used in this paper. If $\bX_1$ is the binary data matrix, then we assume that there is a low-dimensional deterministic structure $\boldsymbol{\Theta_1} (I \times J_1)$ underlying $\bX_1$ and the elements of $\bX_1$ follow a Bernoulli distribution with parameters $\phi(\theta_{1ij})$, thus $x_{1ij} \sim B(\phi(\theta_{1ij}))$. The function $\phi(.)$ can be taken as the logit link $\phi(\theta)=(1+exp(-\theta))^{-1}$ and $x_{1ij}$, $\theta_{1ij}$ are the $ij^{th}$ elements of $\bX_1$ and $\boldsymbol{\Theta_1}$, respectively. The $\boldsymbol{\Theta_1}$ is now assumed to be equal to $\boldsymbol{1} \boldsymbol{\mu}_1^T + \bZ \bA_1$ where $\boldsymbol{\mu}_1$ represent the off-set term, $\bZ$ the common scores and $\bA_1$ the loadings for the binary data.\\

\noindent The quantitative measurements $\bX_2$ are assumed to follow the model $\bX_2=\boldsymbol{1} \boldsymbol{\mu}_2^T + \bZ \bA_2 + \bE$ where the elements $e_{ij}$ of $\bE$ are normally distributed with mean 0 and variance $\sigma^2$. The matrix $\bA_2$ contains the loadings of the quantitative data set; $\bZ$ are again the common scores and the constraints $\bZ^T\bZ=I \bI_R$ and $\boldsymbol{1}^T\bZ=0$ are imposed for identifiability. The shared information between $\bX_1$ and $\bX_2$ is assumed to be represented fully by the common latent variables $\bZ$. Thus $\bX_1$ and $\bX_2$ are stochastically independent given these latent variables and the negative log-likelihoods of both parts can be summed:
\begin{gather}
  f_1(\boldsymbol{\Theta_1})  =  -\sum_i^I \sum_j^{J_1}[x_{1ij}log(\phi(\theta_{1ij}))+(1-x_{1ij})log(1-\phi(\theta_{1ij})] \\ \nonumber
  f_2(\boldsymbol{\Theta_2},\sigma^2)  =  \frac{1}{2\sigma^2}\|\bX_2-\boldsymbol{\Theta_2}\|_F^2 + \frac{1}{2}log(2\pi\sigma^2) \\ \nonumber
  f(\boldsymbol{\Theta_1},\boldsymbol{\Theta_2},\sigma^2) = f_1(\boldsymbol{\Theta_1})+f_2(\boldsymbol{\Theta_2},\sigma^2)
\end{gather}
and minimized simultaneously. This requires some extra constraints; details are given elsewhere \citep{Song2018}.\\

\section{Practical issues and examples}

\subsection{Genomics example}

The genomics example is from the field of cancer research and the data are obtained from the Genomics in Drugs Sensitivity in Cancer from the Sanger Institute (http://www.cancerrxgene.org/). Briefly, this repository consists of measurements performed on cell lines pertaining to different types of cancer. We used the copy number aberration (CNA) and gene-expression data of the cell lines related to breast cancer (BRCA), lung cancer (LUAD) and skin cancer (SKCM). After selecting the samples which had values for all these types of cancer we filtered the gene-expression data by selecting the 1000 variables with the highest variance across the samples. The CNA data contains amplifications and losses of DNA-regions as compared to the average copy numbers in the population. Both amplifications and losses are encoded as ones indicating deviances. The zeros in the CNA data indicate a normal diploid copy number. This provides us with $I=160$ samples; $J_1=410$ binary values for the CNA data and $J_2=1000$ variables for the gene-expression data.\\

\noindent For the representation approach we built a three-way array of size $160 \times 160 \times (410+1000)$ and performed an IDIOMIX analysis. For the binary part, this array contains the slabs $\bS_j$ according to Eqn. \ref{eRMNom1} and for the gene-expression part, the slabs $\bS_j$ are defined by the outer products of the samples in the gene-expression data after auto-scaling the columns of that data. The optimal scaling result are obtained by auto-scaling both raw data sets and subsequently perform an (OS-)SCA on the concatenated data $\bX_{sc}=[\bX_{1sc} | \bX_{2sc}]$. The final way of fusing the two data sets is by using the GSCA model.\\

\noindent The amounts of explained variations are shown in Table \ref{Table1} which contains a lot of information and should be interpreted with care.
\begin{table}[h!]
\footnotesize
\centering
\begin{tabular}{|c|ccc|ccc|ccc|c|}
 \hline
 Method  & IDIOMIX & & & OS-SCA & & & GSCA & & & PCA \\ \hline
 Data type & Binary & Quant & Total & Binary & Quant & Total & Binary & Quant & Total & Quant \\ \hline
 SC1 & 0.06 & 9.32 & 6.60 & 13.65 & 22.15 & 16.14 & 63.58 & 22.11 & 23.64 & 22.15 \\
 SC2 & 0.03 & 2.38 & 1.69 & 5.66 & 10.17 & 7.48 & 15.72 & 10.19 & 10.39 & 10.17 \\
 SC3 & 3.73 & 0.01 & 1.10 & 4.99 & 4.52 & 4.35 & 6.17 & 4.48 & 4.54 & 4.52 \\ \hline
 Cum & 3.82 & 11.70 & 9.39 & 24.29 & 36.84 & 27.96 & 85.47 & 36.77 & 38.58 & 36.84 \\ \hline
\end{tabular}
\caption{\label{Table1}\footnotesize Variances explained by the various methods. SC is the abbreviation of simultaneous component. For more explanation, see main text.}
\label{Table1}
\end{table}
First, for IDIOMAX, OS-SCA and the quantitative part of the GSCA model the explained variation is calculated using sums-of-squares. This is not the case for the binary part of GSCA (for details, see \cite{Song2018}). Second, IDIOMAX on the one hand and OS-SCA, GSCA on the other hand are very different types of models, i.e., they use the data directly (OS-SCA, GSCA) or indirectly (IDIOMAX) so a simple comparison of explained sums-of-squares between these types of models is difficult. The final column of the table reports the amounts of explained variation of a regular PCA on the (autoscaled) gene-expression data.\\

\noindent The first observation to make regarding the values in Table \ref{Table1} is that the amounts of explained variations of the PCA model of the gene-expression data is closely followed by the amounts of explained variations in the gene-expression simultaneous components for OS-SCA and GSCA. This means that the data fusion is dominated by the gene-expression block. This is confirmed by plotting the scores of PC1 and PC2 of the PCA on gene-expression against the SC-scores 1 and 2 of OS-SCA and GSCA: these are almost on a straight line (plot not shown). Although the explained variances for IDIOMAX are much lower, the same observation is valid for IDIOMAX: also for this method the first two SC-scores resembles the ones of a PCA on the gene-expression almost perfectly. This dominance of the gene-expression block in the data fusion as reflected in the first two components cannot completely be explained by the differences in block sizes (1000 variables for gene-expression and 410 variables for the CNA block) but is also due to dominant intrinsic systematic patterns in the gene-expression data.\\

\begin{figure}[h!]
 \centerline{\includegraphics*[width=18cm]{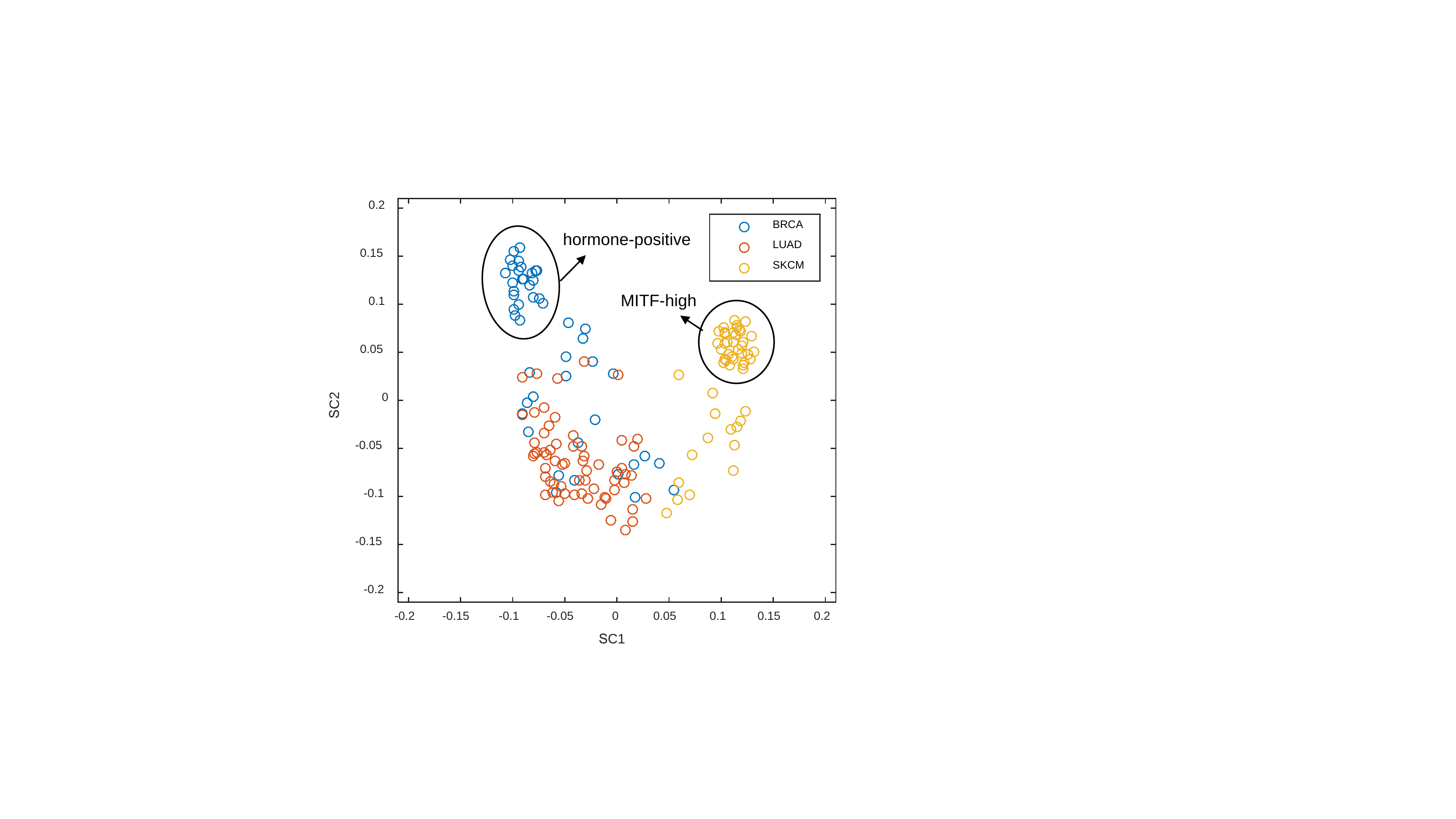}}
 \caption{\footnotesize Scores on SC1 and SC2 for the GSCA model (see text).}
 \label{FigurePMPC1PC2}
\end{figure}

\noindent To get a feeling for what is represented in the first two SCs (that are virtually identical across the three methods), we show the scores for the GSCA method on SC1 and SC2 in Figure \ref{FigurePMPC1PC2}.
The scores show a clear separation in cancer types with specific sub-clusters for hormone-positive breast cancer (within the BRCA-group) and MITF-high melanoma (in the SKCM group) (for a more elaborate interpretation see \cite{Song2018}).\\

\begin{figure}[h!]
 \centerline{\includegraphics*[width=15cm]{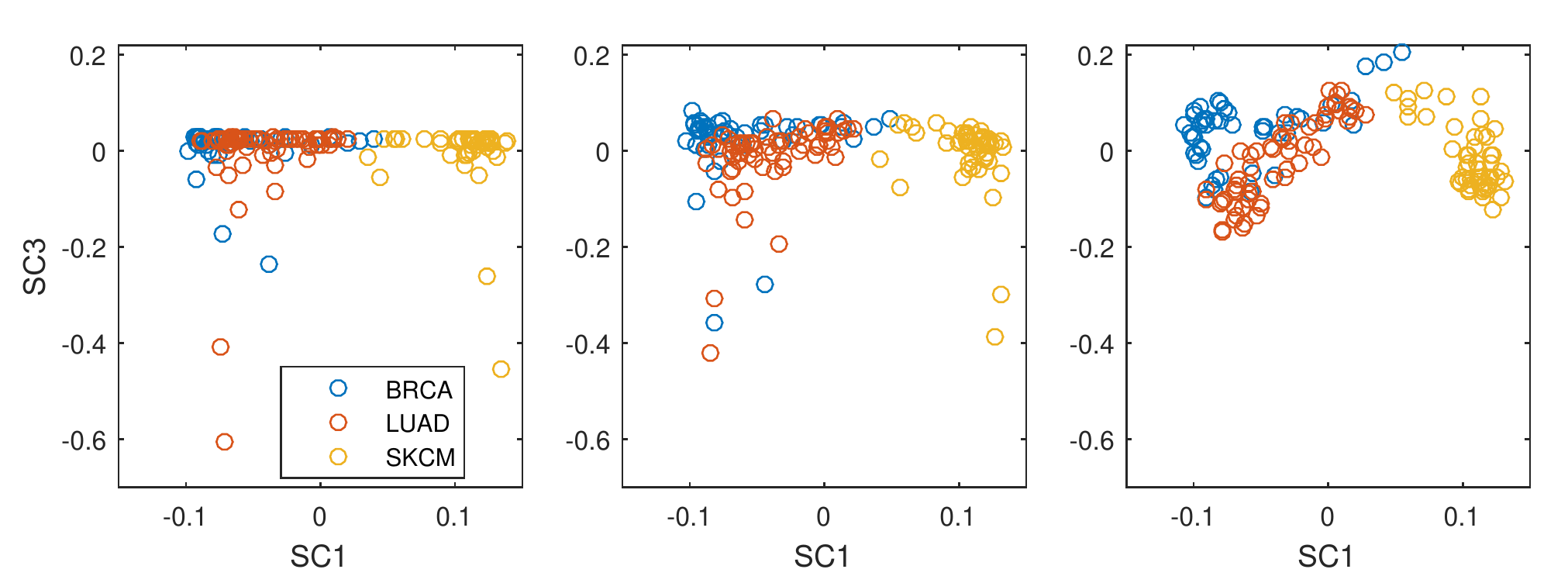}}
 \caption{\footnotesize Score plots of SC1 versus SC3 of in the genomics example using all models (see text). Legend: left: IDIOMIX; middle: OS-SCA; right: GSCA.}
 \label{FigureGenPC1PC3}
\end{figure}

\begin{figure}[h!]
 \centerline{\includegraphics*[width=18cm]{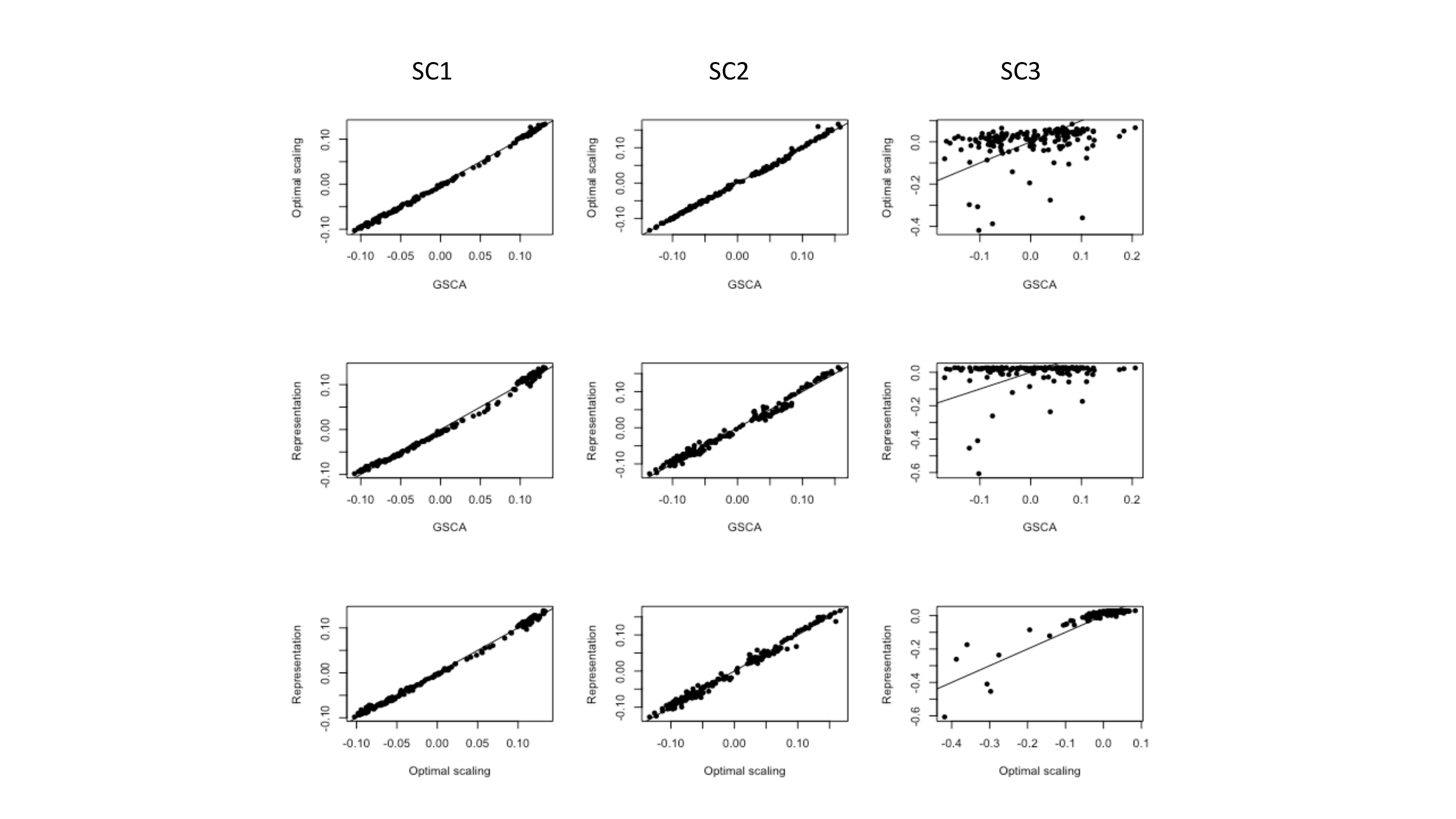}}
 \caption{\footnotesize Score plots of SC1-SC3 for all fusion methods. Optimal scaling is OS-SCA; Representation is IDIOMIX.}
 \label{PC1PC2PC3}
\end{figure}

\begin{figure}[h!]
 \centerline{\includegraphics*[width=18cm]{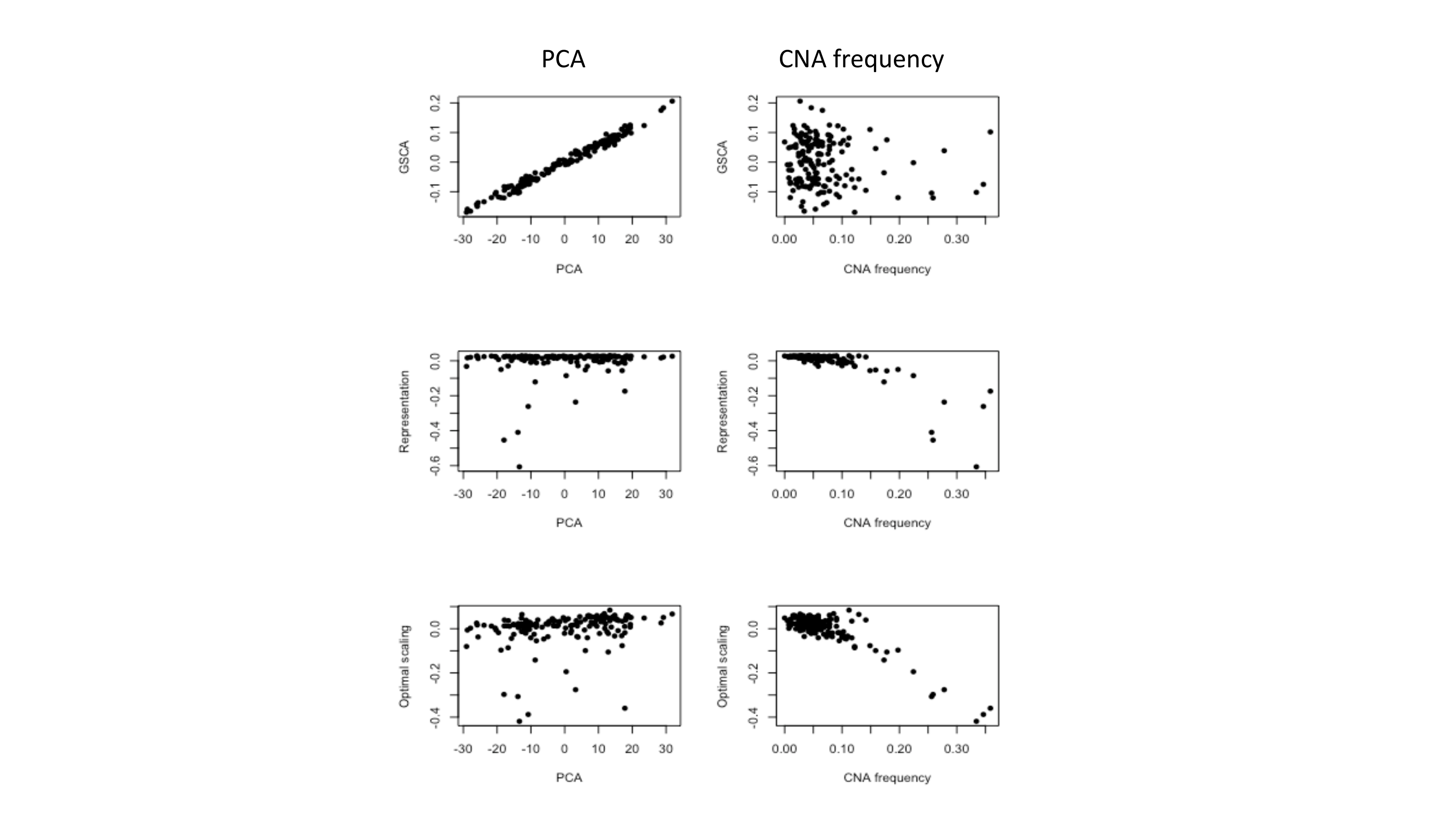}}
 \caption{\footnotesize Left panels: score plots of PC3 of PCA on gene-expression data (x-axis) compared to the SC3 from the fusion results (y-axis); optimal scaling is OS-SCA, representation is IDIOMAX. Right panels: scores of SC3 of all methods compared to CNA frequency.}
 \label{PCACNAFreq}
\end{figure}

\noindent Whereas the three approaches give similar results for the first two simultaneous components, qualitative differences can be seen in SC3. This is especially apparent in Table \ref{Table1} where the third component for IDIOMIX is now dominated by the CNA data. This is visualized in Figure \ref{FigureGenPC1PC3} which shows the score plots of SC1 versus SC3 for all methods which are clearly different.
To further confirm this, the scores of the different methods for the three different components were plotted against each other (see Figure \ref{PC1PC2PC3}) and this confirms that indeed the first two SCs are very similar for all methods, but that SC3 shows differences where GSCA is the most deviating.
To shed some light on this deviating behavior, we plotted the scores of a PCA on the gene-expression data against the SC-scores of the fusion methods for the third component, see Figure \ref{PCACNAFreq}.
The left panels in this figure show that SC3 from GSCA is very similar to the PC3 of a PCA on the gene-expression data alone (see also again Table \ref{Table1}). The same does not hold true for the other methods. The CNA values are available for each sample and thus the scores on the fusion SC3 can be plotted against the frequency at which such an aberration occurs (number of ones divided by the total). From the right panels of Figure \ref{PCACNAFreq}, it then becomes clear that SC3 of IDIOMAX and OS-SCA are mostly picking up the differences in frequencies, contrary to the GSCA-SC3 scores.\\

\begin{figure}[h!]
 \centerline{\includegraphics*[width=18cm]{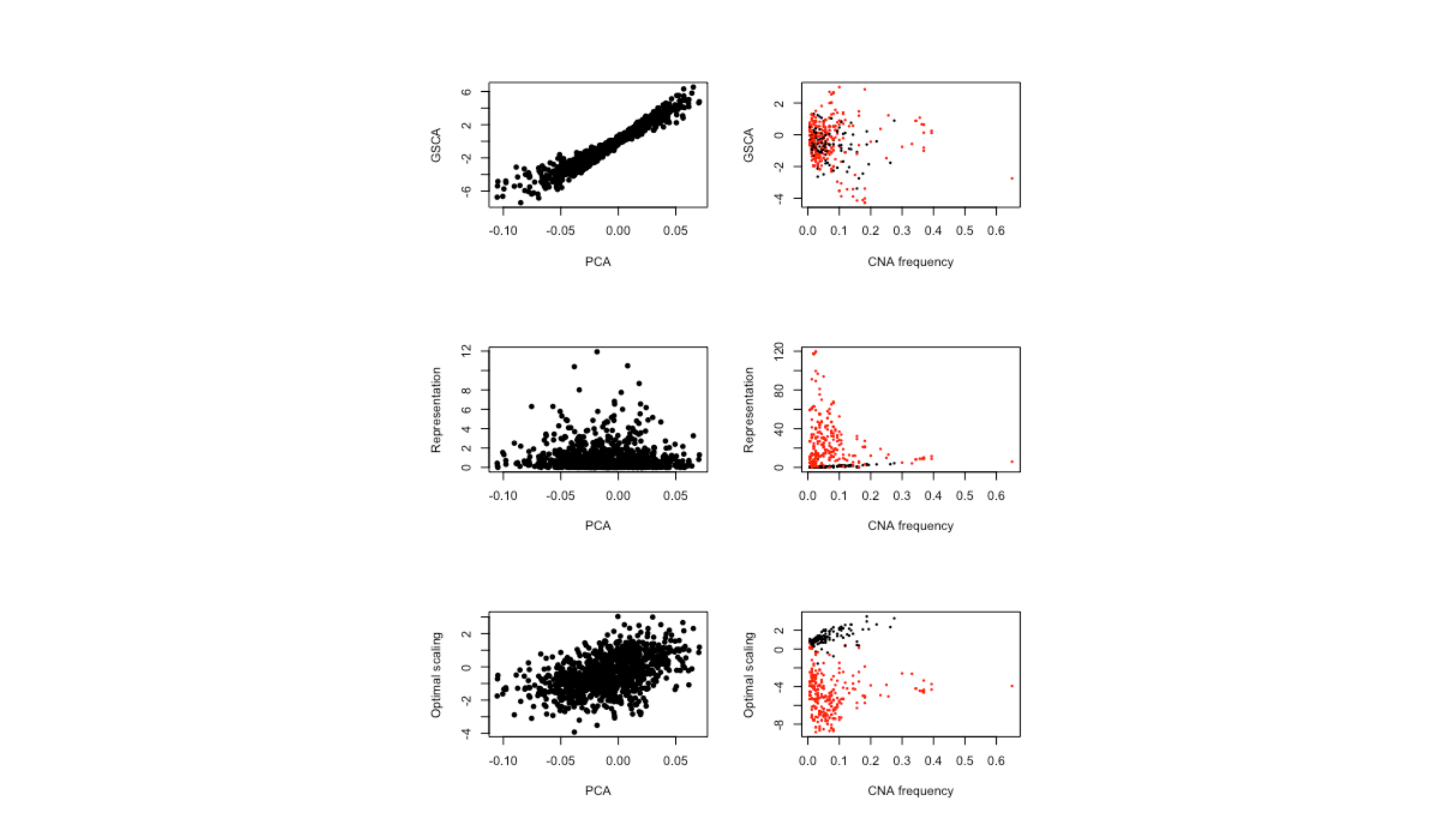}}
 \caption{\footnotesize Left panels: PC3 loadings of PCA on gene-expression data (x-axis) compared to the SC3 loadings from the fusion results (y-axis); optimal scaling is OS-SCA, representation is IDIOMIX. Right panels: loadings of SC3 of all fusion methods Amplifications (black) and Losses (red).}
 \label{PC3GainsLoss}
\end{figure}

\begin{figure}[h!]
 \centerline{\includegraphics*[width=15cm]{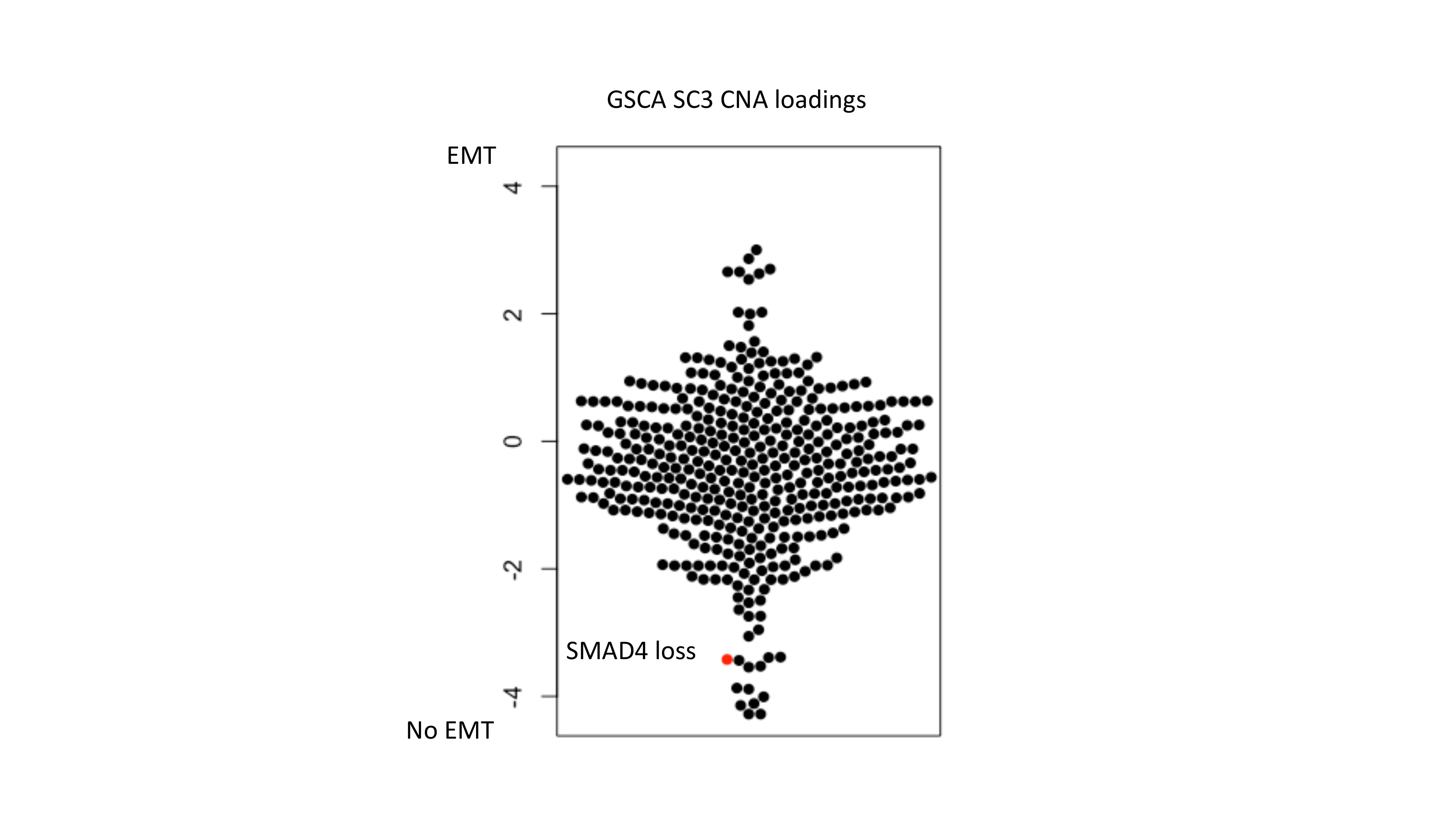}}
 \caption{\footnotesize SC3 loadings of GSCA. In red: SMAD4loss (see text).}
 \label{PC3EMT}
\end{figure}

\noindent A similar comparison can be made for the loadings, see Figure \ref{PC3GainsLoss}. The left panels show the PC3 loadings from gene-expression using PCA and the fusion methods. In the right panels the fusion loadings are plotted against CNA frequencies (now across DNA-positions) and those show no correlation. As explained earlier, the aberrations can either be amplifications or losses and those are clearly picked up by the loadings of IDIOMAX and OS-SCA.\\
\noindent To interpret the GSCA-loadings, these loadings were subjected to a Gene Set Enrichment Analysis (GSEA). This resulted in a highly significant enrichment for epithelial-mesenchymal transition (EMT), a process undergone by tumor cells frequently associated with invasion of surrounding tissues and subsequent metastases. The largest positive loading on GSCA-PC3 for the gene-expression is ZEB1, a transcription factor associated with EMT. A plot of the loadings of the CNA data is shown in Figure \ref{PC3EMT} and one of the loadings identifies SMAD4 loss as an important factor. SMAD4 is required for TGF-$\beta$ driven EMT which confirms the finding that the GSCA gene-expression loadings are strongly enriched for EMT \citep{Tian2009}.\\

\noindent Summarizing, IDIOMAX and OS-SCA are very similar for the whole analysis. For the first two SCs, also the GSCA resembles the other approaches. The difference of GSCA is in the third SC. It seems that GSCA is focussing more on the gene-expression data; whereas IDIOMAX and OS-SCA pick up specific aspects of the CNA data in this third SC. The results of the GSCA-SC3 are biologically relevant; this is less the case for SC3 of the other approaches. It may be that GSCA is focussing more on the common variation between the two data sets and is less influenced by the distinctive parts \citep{Smilde2017}. This needs further exploration in a follow-up paper.

\section{Discussion}
\label{Discussion}

In this paper we have described and compared three methods of fusing data of different measurement scales. We used the example of quantitative and binary data, but all methods can also deal with ordinal data. For the example, it appears that IDIOMAX and OS-SCA give very similar results whereas GSCA is different. One of the reasons may be that the methods deal differently with common and distinct parts of the data.\\

\noindent All methods have meta-parameters, that is, prior choices have to be made. For IDIOMAX, this is the type of representation to select; for OS-SCA it is the type of restrictions to apply; for GSCA it is the distribution to assume for the separate data sets. All methods also require selecting the complexity of the models, i.e., the number of components. The selection of all these meta-parameters will, in practice, be made based on a mixture of domain knowledge and validation, such as cross-validation or scree-tests for selecting model complexity.\\

\noindent We hesitate in giving recommendations regarding which method to use for a particular application. First, the example of this paper concerns an exploratory study for which it is always difficult to judge the relative merits of the methods. Secondly, the cultural background of the investigator plays a role. In data analysis and chemometrics, the culture is to avoid distributional assumptions and have a more data analytic approach, thus resulting in a preference for IDIOMAX or OS-SCA. In statistics and, to some extent, in bioinformatics there is more a tendency to go for parametric modes, hence, GSCA in our context. Thirdly, these methods have not yet been used to a large extent by researchers, hence, experience on their behavior upon which a recommendation can be based is lacking.\\

\noindent In terms of ease of use, we have a slight preference for IDIOMAX. Once the representation matrices are built, standard three-way analysis software can be used to fit the models. There is also software available for OS-SCA and GSCA, but this software is more difficult to implement.\\

\noindent There remain open issues to be investigated. Some of the more prominent ones is to understand the behavior of the methods regarding common, distinct and local components in fusing data sets. Little has been done in this field regarding data of different measurement scales.

\section{Appendix}
\label{Appendix}

\subsection{Optimal scaling of binary data equals analyzing standardized data}

The fact that optimal scaling of binary data equals the analysis of standardized data can be shown as follows. Suppose that a binary vector has $n_0$ values of zero, $n_1$ values of one and $n$ values in total, and $a$ is the optimal scaled value for the zeros and $b$ for the ones. In optimal scaling, the optimal scaled variables need to get some kind of normalization. A common set of choices (see \citep{Gifi1990}) is to make sure that the scaled values have mean zero and variance one. This leads to the following two equations:
\begin{gather}\label{eProof1}
  n_0a+n_1b=0 \\ \nonumber
  n_0a^2+n_1b^2=n \nonumber
\end{gather}
and these equations can be solved for $a$ and $b$ since $n_0, n_1$ and $n$ are known. This gives two values for $a$; one positive and one negative. The values of $b$ follow automatically with the opposite sign. Hence, both solutions are practically equal.

\subsection{Examples of representation matrices for ratio- and interval-scaled data}
\label{Examples ratio scaled}

We will illustrate some ideas of representation matrices using a small example of an $(4 \times 2)$ matrix $\bX=[\bx_1|\bx_2]$:
\begin{eqnarray}
 \bX
 =
 \left(
 \begin{array}{cc}
 2 & 9 \\
 4 & 9 \\
 6 & 10 \\
 8 & 12 \\
 \end{array}
 \right)
\label{esmallX}
\end{eqnarray}
and the standardized version of this is
\begin{eqnarray}
 \bX_s
 =
 \left(
 \begin{array}{cc}
 -0.671 & -0.408 \\
 -0.224 & -0.408 \\
 0.224 & 0 \\
 0.671 & 0.816 \\
 \end{array}
 \right)
\label{esmallXs}
\end{eqnarray}
where indeed ${\bx^T}\!_{s1}{\bx}_{s1}=1$, ${\bx^T}\!_{s2}\bx_{s2}=1$ and ${\bx^T}\!_{s1}\bx_{s2}=0.913$ the latter being the correlation between $\bx_1$ and $\bx_2$. The square representation using Eqn. \ref{eRM1} on $\bx_1$ gives
\begin{eqnarray}
 \widetilde{\bS}_1
 =
 \left(
 \begin{array}{cccc}
 0 & -2 & -4 & -6\\
 2 & 0 & -2 & -4\\
 4 & 2 & 0 &  -2\\
 6 & 4 & 2 &  0\\
 \end{array}
 \right)
\label{esmallS1}
\end{eqnarray}
which is skew-symmetric (${\widetilde{\bS}^T}_1=-\widetilde{\bS}_1$) and contains all the differences between the elements of $\bx_1$. The standardized version of $\widetilde{\bS}_1$ is
\begin{eqnarray}
 \bS_1
 =
 \left(
 \begin{array}{cccc}
 0 & -0.158 & -0.316 & -0.474\\
 0.158 & 0 & -0.158 & -0.316\\
 0.316 & 0.158 & 0 &  -0.158\\
 0.474 & 0.316 & 0.158 &  0\\
 \end{array}
 \right)
\label{esmallS1s}
\end{eqnarray}
and a similar matrix can be made for $\bx_2$. Then using Eqn. \ref{eAss2} on the pairs $(\bS_1,\bS_1)$ and $(\bS_2,\bS_2)$ gives a value of one; and on the pair $(\bS_1,\bS_2)$ gives 0.913, which is the Pearson's correlation again.\\

\noindent Alternative square representations of $\bx_{s1}$ and $\bx_{s2}$ are
\begin{eqnarray}
 \bS_{A1}
 =\bx_{s1} {\bx^T}\!_{s1}=
 \left(
 \begin{array}{cccc}
 0.45 & 0.15 & -0.15 & -0.45\\
 0.15 & 0.05 & -0.05 & -0.15\\
 -0.15 & -0.05 & 0.05 & 0.15\\
 -0.45 & -0.15 & 0.15 & 0.45\\
 \end{array}
 \right)
\label{esmallSA1}
\end{eqnarray}
and
\begin{eqnarray}
 \bS_{A2}
 =\bx_{s2} {\bx^T}\!_{s2}=
 \left(
 \begin{array}{cccc}
 0.167 & 0.167 & 0 & -0.333\\
 0.167 & 0.167 & 0 & -0.333\\
 0 & 0 & 0 & 0\\
 -0.333 & -0.333 & 0 & 0.667\\
 \end{array}
 \right)
\label{esmallSA2}
\end{eqnarray}
and using Eqn. \ref{eAss2} on $\bS_{A1}$ and $\bS_{A2}$ gives 0.833 which is the squared Pearson's correlation between the original variables.

\subsection{Examples of representation matrices for nominal data}
\label{Examples nominal data}

We will illustrate some ideas on representing nominal data using two categorical variables $\bx_1$ and $\bx_2$. The first variable contains four categories encoded as A,B,C,D and reads $\bx_1=(A,B,A,C,D,C,B,D)^T$; the second variable has three categories encoded as I,II,III and reads $\bx_2=(I,II,II,I,III,III,I,II)^T$ where the roman capitals are used to show that the two variables encode different types of categories. The indicator matrices are now
\begin{eqnarray}
 \bG_1
 =
 \left(
 \begin{array}{cccc}
 1 & 0 & 0 & 0\\
 0 & 1 & 0 & 0\\
 1 & 0 & 0 & 0\\
 0 & 0 & 1 & 0\\
 0 & 0 & 0 & 1\\
 0 & 0 & 1 & 0\\
 0 & 1 & 0 & 0\\
 0 & 0 & 0 & 1\\
 \end{array}
 \right)
\label{eG1nom}
\end{eqnarray}
and
\begin{eqnarray}
 \bG_2
 =
 \left(
 \begin{array}{ccc}
 1 & 0 & 0 \\
 0 & 1 & 0 \\
 0 & 1 & 0 \\
 1 & 0 & 0 \\
 0 & 0 & 1 \\
 0 & 0 & 1 \\
 1 & 0 & 0 \\
 0 & 1 & 0 \\
 \end{array}
 \right)
\label{eG2nom}
\end{eqnarray}
and a special feature of this kind of data becomes present namely that some objects have exactly the same rows in $\bG_1$ (and similarly in $\bG_2$). Moreover, the matrices show closure (${\bG^T}\!_1 \mathbf{1}=\mathbf{1},{\bG^T}\!_2 \mathbf{1}=\mathbf{1}$). The marginal frequencies are collected in
\begin{eqnarray}
 \bD_1
 =
 \left(
 \begin{array}{cccc}
 2 & 0 & 0 & 0\\
 0 & 2 & 0 & 0\\
 0 & 0 & 2 & 0\\
 0 & 0 & 0 & 2\\
 \end{array}
 \right)
 =\bG^T_1\bG_1
\label{eD1nom}
\end{eqnarray}
and
\begin{eqnarray}
 \bD_2
 =
 \left(
 \begin{array}{ccc}
 3 & 0 & 0 \\
 0 & 3 & 0 \\
 0 & 0 & 2 \\
 \end{array}
 \right)
 = \bG^T_2\bG_2
\label{eD2nom}
\end{eqnarray}
with obvious properties.\\

\noindent Simple representations of these variables are now
\begin{eqnarray}
 \bS_{1s}
 =
 \left(
 \begin{array}{cccccccc}
 1 & 0 & 1 & 0 & 0 & 0 & 0 & 0\\
 0 & 1 & 0 & 0 & 0 & 0 & 1 & 0\\
 1 & 0 & 1 & 0 & 0 & 0 & 0 & 0\\
 0 & 0 & 0 & 1 & 0 & 1 & 0 & 0\\
 1 & 0 & 1 & 0 & 0 & 0 & 0 & 0\\
 0 & 0 & 0 & 0 & 1 & 0 & 0 & 1\\
 0 & 0 & 0 & 1 & 0 & 1 & 0 & 0\\
 0 & 0 & 0 & 0 & 1 & 0 & 0 & 1\\
 \end{array}
 \right)
 =\bG_1\bG^T_1
\label{eS1simple}
\end{eqnarray}
and
\begin{eqnarray}
 \bS_{2s}
 =
 \left(
 \begin{array}{cccccccc}
 1 & 0 & 0 & 1 & 0 & 0 & 1 & 0\\
 0 & 1 & 1 & 0 & 0 & 0 & 0 & 1\\
 0 & 1 & 1 & 0 & 0 & 0 & 0 & 1\\
 1 & 0 & 0 & 1 & 0 & 0 & 1 & 0\\
 0 & 0 & 0 & 0 & 1 & 1 & 0 & 0\\
 0 & 0 & 0 & 0 & 1 & 1 & 0 & 0\\
 1 & 0 & 0 & 1 & 0 & 0 & 1 & 0\\
 0 & 1 & 1 & 0 & 0 & 0 & 0 & 1\\
 \end{array}
 \right)
 =\bG_2\bG^T_2
\label{eS2simple}
\end{eqnarray}
and these representations are encoding which objects have equal categories in the variables. The more complex representations (according to Eqn \ref{eRMNom1}) are now
\begin{eqnarray}
 \bS_{1c}
 =
 \left(
 \begin{array}{cccccccc}
 0.375 & -0.125 & 0.375 & -0.125 & -0.125 & -0.125 & -0.125 & -0.125\\
 -0.125 & 0.375 & -0.125 & -0.125 & -0.125 & -0.125 & 0.375 & -0.125\\
 0.375 & -0.125 & 0.375 & -0.125 & -0.125 & -0.125 & -0.125 & -0.125\\
 -0.125 & -0.125 & -0.125 & 0.375 & -0.125 & 0.375 & -0.125 & -0.125\\
 -0.125 & -0.125 & -0.125 & -0.125 & 0.375 & -0.125 & -0.125 & 0.375\\
 -0.125 & -0.125 & -0.125 & 0.375 & -0.125 & 0.375 & -0.125 & -0.125\\
 -0.125 & 0.375 & -0.125 & -0.125 & -0.125 & -0.125 & 0.375 & -0.125\\
 -0.125 & -0.125 & -0.125 & -0.125 & 0.375 & -0.125 & -0.125 & 0.375\\
 \end{array}
 \right)
\label{eS1complex}
\end{eqnarray}
and
\begin{eqnarray}
 \bS_{2c}
 =
 \left(
 \begin{array}{cccccccc}
 0.208 & -0.125 & -0.125 & 0.208 & -0.125 & -0.125 & 0.208 & -0.125\\
 -0.125 & 0.208 & 0.208 & -0.125 & -0.125 & -0.125 & -0.125 & 0.208\\
 -0.125 & 0.208 & 0.208 & -0.125 & -0.125 & -0.125 & -0.125 & 0.208\\
 0.208 & -0.125 & -0.125 & 0.208 & -0.125 & -0.125 & 0.208 & -0.125\\
 -0.125 & -0.125 & -0.125 & -0.125 & 0.375 & 0.375 & -0.125 & -0.125\\
 -0.125 & -0.125 & -0.125 & -0.125 & 0.375 & 0.375 & -0.125 & -0.125\\
 0.208 & -0.125 & -0.125 & 0.208 & -0.125 & -0.125 & 0.208 & -0.125\\
 -0.125 & 0.208 & 0.208 & -0.125 & -0.125 & -0.125 & -0.125 & 0.208\\
 \end{array}
 \right)
\label{eS2complex}
\end{eqnarray}
which are indeed double centered and standardized. Using Eqn. \ref{eAss2} on the matrices $\bS_{1c}$ and $\bS_{2c}$ gives the correlation coefficient $T^2=0.5$.

\subsection{Examples of representation matrices for binary data}
\label{Examples binary data}

As an example for binary data we will use a simple data set consisting of two binary variables $\bx_1$ and $\bx_2$ which are columns of
\begin{eqnarray}
 \bX
 =
 \left(
 \begin{array}{ccc}
 0 & 1 \\
 0 & 1 \\
 1 & 0 \\
 1 & 1 \\
 0 & 1 \\
 1 & 0 \\
 0 & 1 \\
 0 & 0 \\
 \end{array}
 \right)
\label{eG1bin}
\end{eqnarray}
with indicator matrices
\begin{eqnarray}
 \bG_1
 =
 \left(
 \begin{array}{ccc}
 1 & 0 \\
 1 & 0 \\
 0 & 1 \\
 0 & 1 \\
 1 & 0 \\
 0 & 1 \\
 1 & 0 \\
 1 & 0 \\
 \end{array}
 \right)
\label{eG1bin}
\end{eqnarray}
and
\begin{eqnarray}
 \bG_2
 =
 \left(
 \begin{array}{ccc}
 0 & 1 \\
 0 & 1 \\
 1 & 0 \\
 0 & 1 \\
 0 & 1 \\
 1 & 0 \\
 0 & 1 \\
 1 & 0 \\
 \end{array}
 \right).
\label{eG2bin}
\end{eqnarray}
A correlation measure between binary variables is the $\phi$-coefficient which is defined as
\begin{equation}\label{ePhi1}
    \frac{n_{11}n_{00}-n_{10}n_{01}}{\sqrt{n_{1.}n_{0.}n_{.0}n_{.1}}}
\end{equation}
where the values $n$ are shown in Table \ref{table2}. For the example, this $\phi$-coefficient equals $-0.4667$ which is also equivalent to the Pearson correlation between $\bx_1$ and $\bx_2$.

\begin{table}[h!]
\centering
\begin{tabular}{|c|cc|c|}
 \hline
         & $x_2=0$   & $x_2=1$    & total \\ \hline
 $x_1=1$ & $n_{11}(1)$ & $n_{10}(2)$  & $n_{1.}(3)$ \\
 $x_1=0$ & $n_{01}(4)$ & $n_{00}(1)$  & $n_{0.}(5)$ \\ \hline
 total   & $n_{.1}(5)$ & $n_{.0}(3)$ & $n(8)$ \\ \hline
\end{tabular}
\caption{\label{Table2}\footnotesize Calculation of the $\phi$-coefficient (between brackets the values of the example).}
\label{table2}
\end{table}

\noindent There are two alternative square representations of $\bx_1$ and $\bx_2$. The first uses Eqn. \ref{eRMNom1} based on the indicator matrices and the results are
\begin{eqnarray}
 \bS_{1c}
 =
 \left(
 \begin{array}{cccccccc}
 0.075 &  0.075 & -0.125 & -0.125 &  0.075 & -0.125 & 0.075 & 0.075 \\
 0.075 & 0.075 & -0.125 & -0.125 &  0.075 & -0.125 &  0.075 & 0.075 \\
 -0.125 & -0.125 & 0.208 & 0.208 & -0.125 & 0.208 & -0.125 & -0.125 \\
 -0.125 & -0.125 & 0.208 & 0.208 & -0.125 & 0.208 & -0.125 & -0.125 \\
  0.075 & 0.075 & -0.125 & -0.125 & 0.075 & -0.125 & 0.075 & 0.075 \\
 -0.125 & -0.125 & 0.208 & 0.208 & -0.125 & 0.208 & -0.125 & -0.125 \\
  0.075 & 0.075 & -0.125 & -0.125 & 0.075 & -0.125 & 0.075 & 0.075 \\
  0.075 & 0.075 & -0.125 & -0.125 & 0.075 & -0.125 & 0.075 & 0.075 \\
 \end{array}
 \right)
\label{eS1Nomcomplex}
\end{eqnarray}
and
\begin{eqnarray}
 \bS_{2c}
 =
 \left(
 \begin{array}{cccccccc}
0.075 & 0.075 & -0.125 & 0.075 & 0.075 & -0.125 & 0.075 & -0.125 \\
0.075 & 0.075 & -0.125 & 0.075 & 0.075 & -0.125 & 0.075 & -0.125 \\
-0.125 & -0.125 & 0.208 & -0.125 & -0.125 & 0.208 & -0.125 & 0.208 \\
 0.075 & 0.075 & -0.125 &  0.075 &  0.075 & -0.125 &  0.075 & -0.125 \\
 0.075 & 0.075 & -0.125 &  0.075 &  0.075 & -0.125 &  0.075 & -0.125 \\
 -0.125 & -0.125 & 0.208 & -0.125 & -0.125 & 0.208 & -0.125 & 0.208 \\
 0.075 & 0.075 & -0.125 & 0.075 & 0.075 & -0.125 &  0.075 & -0.125 \\
 -0.125 & -0.125 & 0.208 & -0.125 & -0.125 & 0.208 & -0.125 & 0.208\\
 \end{array}
 \right)
\label{eS2Nomcomplex}
\end{eqnarray}
and when these are used in Eqn. \ref{eAss2} the result is $0.2178$ which is the square of the $\phi$-coefficient.\\

\noindent The other representations are based on the standardized $x$-variables $\bz_1$ and $\bz_2$ (with ${\bz^T}\!_1 \bz_2=-0.4667$). It now holds that $\bJ \bG_j \bD_j^{-1} \bG_j^T \bJ=\bz_j {\bz^T}\!_j$ and, hence, both representations coincide.

\bibliographystyle{natbib}
\bibliography{main}

\end{document}